%% file: main.tex
\documentclass[conference]{IEEEtran}

\usepackage{cite}
\usepackage{amsmath,amssymb,amsfonts}
\usepackage{graphicx}
\usepackage{xcolor}
\usepackage{booktabs}
\usepackage{tikz}
\usetikzlibrary{tikzmark}
\usepackage{multirow}
\usepackage[hidelinks]{hyperref}

\begin{document}

\definecolor{calloutcolor}{HTML}{000000}
\tikzset{mycircled/.style={circle,draw,inner sep=0.1em,line width=0.04em}}
\def\callout#1{\tikzmarknode[mycircled,draw=calloutcolor,fill=calloutcolor]{t1}{\textcolor{white}{#1}}}

\title{Hardware Fingerprinting FTQC\\ via Quantum Decoder Timing}

\author{\IEEEauthorblockN{Friedrich Doku, Jakub Szefer, and Kaitlin N. Smith}
\IEEEauthorblockA{\textit{Northwestern University} \\
Evanston, Illinois, USA \\
friedy@u.northwestern.edu, jakub.szefer@northwestern.edu, kns@northwestern.edu}
}

\maketitle

\begin{abstract}
\input{sections/abstract}
\end{abstract}

\input{sections/introduction}
\input{sections/threat_model}
\input{sections/methodology}
\input{sections/experimental_setup}
\input{sections/evaluation}
\input{sections/discussion}
\input{sections/related_work}
\input{sections/conclusion}
\input{sections/acknowledgment}

\bibliographystyle{plainurl}
\bibliography{references}

\end{document}

%% file: sections/abstract.tex
As the quantum computing field transitions toward Fault-Tolerant Quantum Computing (FTQC), intensive efforts are focused on scaling architectures and realizing active error correction. However, this shift introduces security surfaces that remain largely unexplored. Fault-tolerant quantum computers pair a quantum processor with a classical decoder that sits on the critical path of every syndrome-extraction round. For the first time, this work demonstrates that the \emph{wall-clock time} each decoder takes to process a syndrome measurement and decoding round constitutes a novel, exploitable hardware side channel on physical quantum hardware. Using per-shot decoder timings from three IBM Heron processors collected over a 68-day window, the decode-time distribution alone allows a passive observer to (i)~reconstruct the shot-by-shot detector-firing distribution and estimate the workload's logical error rate $p_L$, (ii)~infer the code distance in use, and (iii)~fingerprint the specific physical device with up to 89\% accuracy (a random guess is 33\%), with a pooled two-sample Kolmogorov--Smirnov test confirming the decode-time distributions are statistically distinct. In noisy simulation inspired by public data from Google's 105-qubit Willow processor, decoder timing further distinguishes 9 surface-code patches at different locations on the chip with 81\% accuracy, showing the side channel persists on below-threshold fault-tolerant hardware from a different vendor and code family.

%% file: sections/introduction.tex
\section{Introduction}
\label{sec:intro}

The push toward Fault-Tolerant Quantum Computing (FTQC) is accelerating from theoretical roadmaps into imminent hardware deployments. Spurred by national efforts like the U.S. Department of Energy's recent Quantum Genesis initiative, which targets scientifically relevant fault-tolerant systems with hundreds of logical qubits by 2028~\cite{doe2026quantumgenesis}, the engineering focus is rapidly shifting toward active error correction. Operating these systems requires a classical decoder running in the tight, real-time loop between the quantum processor and the scheduler. Every syndrome extraction round produces measurement bits that a decoder must translate into a correction before the next round begins for real-time decoding. If the decoder falls behind, uncompensated syndrome data accumulates, stalling the logical computation~\cite{Battistel_2023}. Consequently, decoder latency is more than a simple performance metric; it is a load-bearing constraint of the entire quantum stack.

In this work, we show that decoding latency is also a critical information leak. As quantum cloud platforms mature, providers will inevitably abstract away the noisy physical layer, restricting users to purely logical qubits. We demonstrate that decoder timing pierces this abstraction veil. Even when isolated from the physical hardware, an attacker can exploit decoder execution times to passively recover logical error rates and fingerprint the host processor.

\begin{figure}[t]
    \centering
    \includegraphics[width=\linewidth]{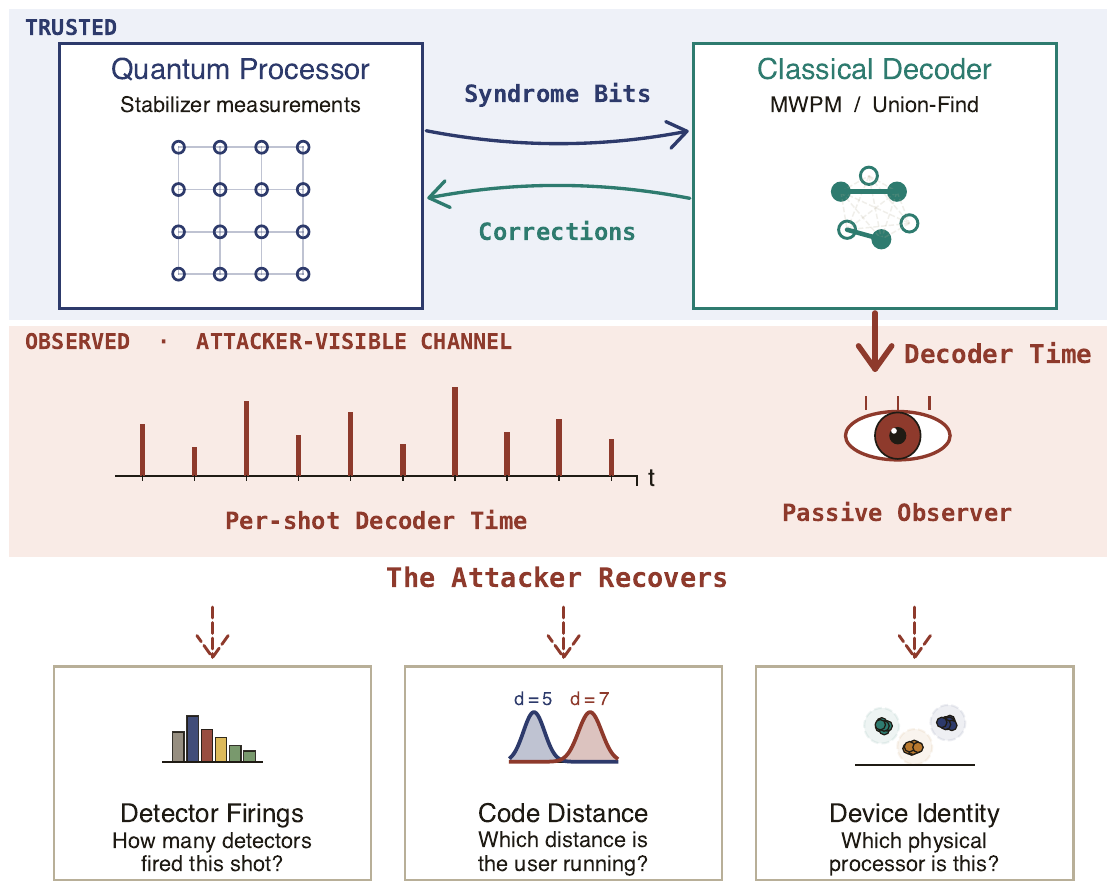}
    \caption{An overview of the timing side-channel on quantum error correction decoders. By measuring the wall-clock time of the classical decoder, a passive observer can extract the detector-firing count, infer the code distance, and fingerprint the underlying physical device.}
    \label{fig:teaser}
\end{figure}

\subsection{The Observation}

A decoder takes longer on a shot that contains more errors to resolve. When the quantum hardware commits more errors, more detectors fire, and the decoder must do more work to pinpoint how noise affected computation. Whether it resolves them by matching pairs on a graph (as in MWPM\cite{mwpmspareblossom}) or by fusing local clusters (as in Union-Find\cite{delfosse2021union}), a denser syndrome directly increases the execution time of the decoding algorithm. The number of detectors that fire on a shot therefore reflects the physical error rates of the measured qubits. Because those rates come from a device-specific mix of gate infidelities, $T_1$/$T_2$ heterogeneity, correlated errors, and readout asymmetry, they leave a characteristic fingerprint of the underlying hardware. Under repeated sampling, this variation manifests as a per-shot decode-time distribution whose \emph{shape} carries a distinct, device-identifying signal. No prior work has demonstrated this leakage channel on quantum hardware.

\subsection{The Stakes}
The consequences of that leak are highest for \emph{device fingerprinting}. Quantum computing is increasingly deployed as a shared cloud resource~\cite{ravi2021quantum}. IBM Quantum\cite{ibmquantum}, AWS Braket\cite{awsbraket}, and Azure Quantum\cite{azurequantum}
all expose their backends to concurrent tenants through runtime APIs. Adopting a public cloud infrastructure as a service (IaaS) model, queues will routinely rotate the physical device that services each job. In this setting, an observer who can identify the underlying device from timing alone gains several capabilities at once. They can confirm co-residency with a target of interest, a prerequisite for cross-tenant attacks. They can defeat provider-level anonymization designed precisely to prevent this reconnaissance.

Similar concerns have motivated a long line of hardware-fingerprinting work on classical shared infrastructure. For instance, FPGAs in cloud computing infrastructures can become targets for information leakages via covert channel communication\cite{covertchannels}. However, establishing these covert channels, requires both the sender and receiver to accurately identify the specific hardware they are communicating across. Reliable device identification is required for cross-tenant data exfiltration, and fingerprinting the machine is a path to achieve device identification.

\subsection{Attacks on Real-hardware}
We executed identical repetition-code memory circuits on three physical IBM Heron processors (\texttt{ibm\_marrakesh}, \texttt{ibm\_fez}, \texttt{ibm\_kingston}) across a 68-day window. We decoded each shot using both MWPM and UF, precisely recording the execution latency of every decoder call.

This timing data enabled:

\begin{enumerate}
    \item \textbf{Detector-count inference.} The per-shot decode-time distribution splits by the number of fired detectors, $n_{\text{firings}}$. A passive observer of decoding time can reconstruct the shot-by-shot firing-count distribution, serving as a proxy for the logical error rate $p_L$, without seeing a single syndrome bit.

    \item \textbf{Code-distance inference.} The latency distributions for different code distances separate distinctly for both decoders. An observer analyzing only the decoder wall-clock times can infer the user's chosen code distance, sharply constraining hypotheses regarding the active workload.
    \item \textbf{Hardware fingerprinting.} We evaluated a $k$-nearest-neighbor classifier on a 14-dimensional feature vector combining the firing-count probability distribution with summary statistics of the decode-time distribution.
\end{enumerate}

We complement these repetition-code attacks with an evaluation on real surface-code data from Google's 105-qubit Willow processor, the first hardware to demonstrate quantum error correction below the surface-code threshold \cite{google2025quantum}.

\smallskip
\noindent \textbf{Contributions.}

\begin{enumerate}
    \item We provide the first demonstration that decoding time is a practical side channel on physical quantum hardware. Our evaluation covers three IBM Heron processors (Sec.~\ref{sec:eval_hw}).
    \item We demonstrate that per-shot decode times permit the reconstruction of the detector-firing count distribution, yielding an effective logical-error-rate estimate without requiring access to syndrome bits (Sec.~\ref{sec:eval_hw}).
    \item We show that decode-time distributions distinguish code distance $d=5$ from $d=7$ for both MWPM and UF, enabling workload inference by a passive observer (Sec.~\ref{sec:eval_hw}).
    \item We construct a 14-dimensional timing feature vector that identifies which of the three physical devices ran a workload with up to 89\% accuracy, where guessing at random would be correct only a third of the time. A pooled two-sample Kolmogorov--Smirnov test confirms that every pair of devices has a distinct timing distribution ($p < 10^{-100}$), with the clearest-separated pair reaching $D = 0.425$ (Sec.~\ref{sec:eval_hw}).
    \item We validate the mechanism on real, below-threshold surface-code data from Google's Willow processor, distinguishing its surface-code patches by decoder timing alone. This shows the information leak is inherent to decoder timing rather than specific to IBM Heron, the repetition code, or current noise levels (Sec.~\ref{sec:sim}).
\end{enumerate}

%% file: sections/threat_model.tex
\section{Threat Model}
\label{sec:threat}

We consider a fault-tolerant quantum architecture consisting of a quantum processor paired with a classical decoder that processes every round of syndrome measurements. We assume the decoder runs an algorithm whose execution time depends on the syndrome it is given, as is the case for the two most widely used families, minimum-weight perfect matching (MWPM) and Union-Find (UF). For every shot, the decoder consumes a syndrome and produces a correction; because denser syndromes take both MWPM and UF longer to resolve, the decoder's \emph{wall-clock execution time} is inherently data-dependent (Section~\ref{sec:eval:detector_count}). This per-shot decode latency is the attack surface, and the only quantity our adversary observes. An attacker could realistically extract these timings through classical channels such as cross-tenant resource contention in shared cloud infrastructure, inadvertently exposed telemetry and debugging metadata, or physical side channels leveraged by an insider. We establish the existence of this channel directly by combining quantum data from a NISQ device (IBM Heron) with a locally-timed decoder (Section~\ref{sec:setup:timer}).

\subsection{Attacker Capabilities}
We assume a \emph{passive, honest-but-curious} adversary whose sole capability is to observe the decoder's per-shot execution time. We treat this observation abstractly: the adversary can measure per-shot decode latency, but we make no assumption about the specific mechanism used. The adversary does not tamper with the computation, inject faults, or interact with the quantum hardware, and it never sees the syndrome bits, the executed circuit, the logical measurement outcomes, or the workload's identity. Every result in this paper is reconstructed solely from decode timing.

\subsection{Attacker Knowledge}
Timing is meaningless in isolation. The adversary therefore holds a modest \emph{reference baseline}: a small set of previously observed decode-time distributions labeled by firing count, code distance, or device. This baseline implicitly encodes the decoder family and code in use, but not the specific workload. As we show in Section~\ref{sec:eval:baselines}, it is cheap to acquire and durable.

\subsection{Attacker Objective}
Using only per-shot decode times and a reference baseline, the adversary aims to (i)~reconstruct the shot-by-shot firing count and thereby estimate the workload's logical error rate $p_L$; (ii)~infer the chosen code distance $d$; and (iii)~fingerprint the specific physical device executing the workload. Together these let the adversary build an unauthorized profile of which device is running which workload, and how well, without ever seeing a single syndrome bit.

%% file: sections/methodology.tex
\section{Methodology}
\label{sec:attack}

\begin{figure}[h]
    \centering
    \includegraphics[width=\linewidth]{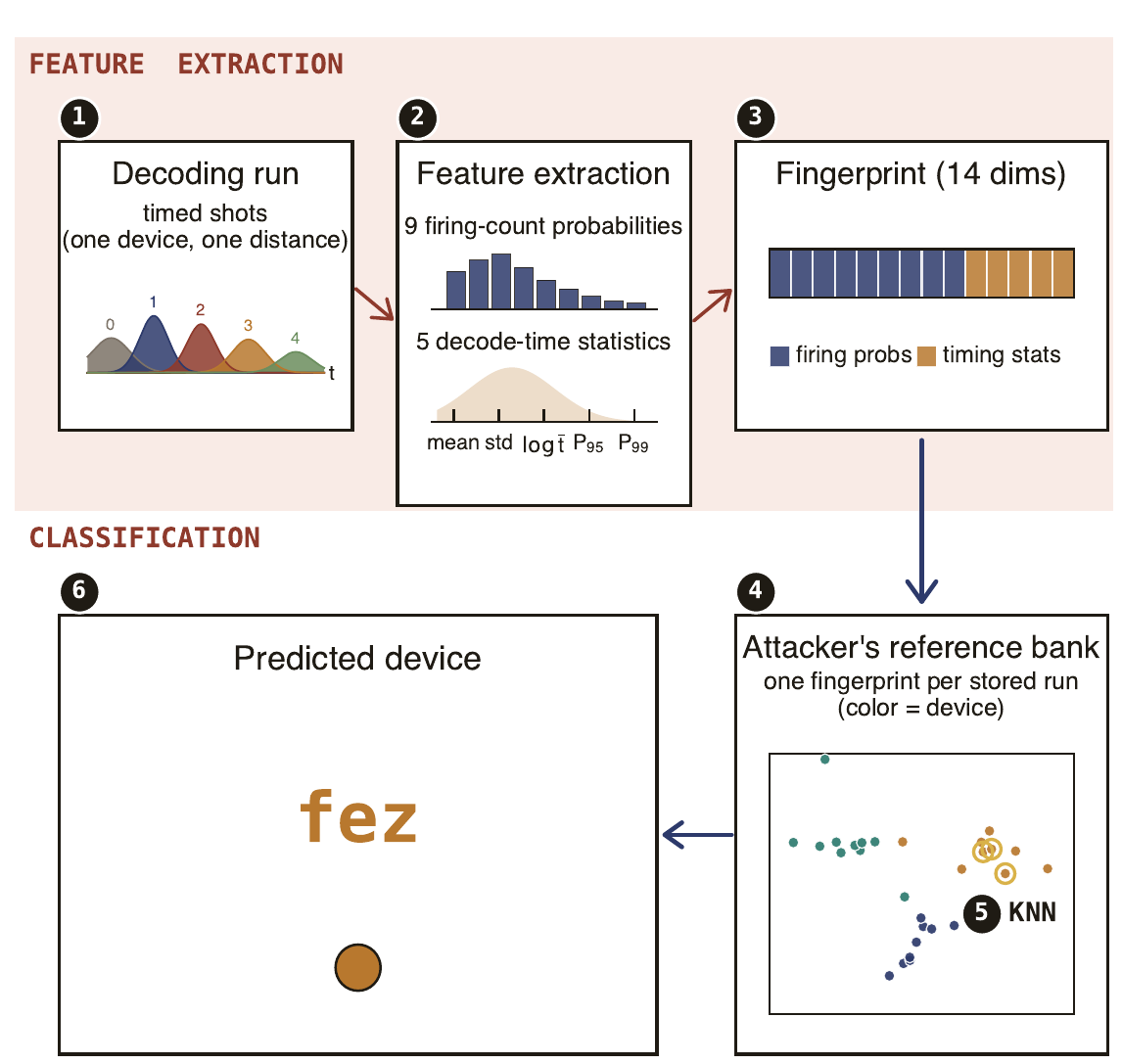}
    \caption{Overview of attack.}
    \label{fig:eval_pipeline}
\end{figure}

We present the full pipeline in Figure~\ref{fig:eval_pipeline}. The adversary observes a single \emph{run} getting the per-shot decode times from one device at it's running code distance \callout{1}. Working only from that run's decode-time distribution, the adversary extracts two complementary families of features. The first one being the distribution of detector-firing counts and the other being a set of summary statistics of the decode times \callout{2}. Put together this forms a 14-dimensional feature vector, which works as a \emph{fingerprint} of the run \callout{3}. To identify the hardware source, the adversary compares this fingerprint against a reference bank of previously collected fingerprints \callout{4}. This data is then fed into a classifier, such as, $k$-nearest-neighbors \callout{5}, which is used to assign the run to a device \callout{6}.

\subsection{Decoder Latency as a Feature Source}
\label{sec:attack:latency}
Every attack in this work depends on a single mechanism: the decoder's runtime. The decoder's runtime increases with the number of detectors that fire. A shot with more fired detectors forces both MWPM and Union-Find to do more work, so its decode takes longer (Section~\ref{sec:eval:detector_count}). Taking the latency per-shot gives us a real-time proxy for the device's noise profile during the shot. Crucially, this mapping is a property of the decoder. It is not dependent on the device, but different devices can cause different decoding latencies due to different noise profiles.

\subsection{Code Distance and Detectors}
\label{sec:attack:features}
The code distance $d$ sets the size of the decoding problem. A larger distance-$d$ surface code or repetition code patch contains more stabilizers, so each round will produce more detectors, which increases the latency of the decoding algorithms overall. Thus, larger $d$ shifts a run's decode-time distribution toward longer times and heavier tails. This makes $d$ visible in timing: two runs of different code distances on the same device produce distinguishable latency distributions, and an observer who has a reference distribution for each candidate distance can infer which one is in use from a modest batch of previously timed shots. Because $d$ is chosen by the operator to hit a target logical failure rate, leaking it also leaks the fidelity regime the workload is aiming for (Section~\ref{sec:eval:distance}).

\subsection{Hardware Fingerprinting}
\label{sec:attack:fingerprint}

All the information required to fingerprint the target device is embedded within its decode-time distribution. Because longer decode times correspond directly to higher detector firing counts, this distribution inherently reflects the hardware's physical noise profile. To extract a reliable fingerprint, we must look beyond a simple average; relying solely on the mean discards critical data because these decode-time distributions are heavily skewed and far from Gaussian.

Formally, we define a \emph{fingerprint} as a 14-dimensional feature vector derived from the run's empirical decode-time distribution (comprising the decoder execution times of 16{,}000 distinct shots). This vector relies on two complementary sets of statistics:

\begin{itemize}
    \item \textbf{Firing-count distribution (9 features):} The probability that a shot fired exactly $k$ detectors ($k \in \{0, 1, \dots, 7\}$), plus a tail bucket for $n_\mathrm{firings} \geq 8$. This maps the physical noise: quieter devices peak at low counts, whereas noisier devices push probability mass into the tail.
    \item \textbf{Decode-time distribution (5 features):} The mean, standard deviation, log-mean, 95th percentile, and 99th percentile of the decode times. These capture how the noise profile dictates runtime, specifically isolating the heavy right tail caused by multi-error corrections.
\end{itemize}

Together, these 14 features encompass the complete hardware fingerprint: the volume of errors generated and the latency to process them. Ablation studies confirm this combination is optimal. Omitting the firing probabilities degrades classification accuracy at $d{=}7$, while omitting the timing percentiles degrades accuracy at $d{=}5$. Busier feature spaces offer no additional predictive power.

\begin{table}[t]
\centering
\small
\caption{The 14 features comprising a fingerprint. All features are extracted from the decode-time distribution.}
\label{tab:eval_fingerprint_features}
\begin{tabular}{@{}llp{4.5cm}@{}}
\toprule
Family & Feature & Description \\
\midrule
\multirow{2}{*}{\shortstack[l]{Firing counts\\(9 features)}}
  & $P(n_\mathrm{firings}{=}k)$ & Fraction of shots firing exactly $k$ detectors, for $k \in \{0, \dots, 7\}$ \\
\addlinespace
  & $P(n_\mathrm{firings}{\geq}8)$ & Fraction of shots firing 8 or more detectors \\
\midrule
\multirow{5}{*}{\shortstack[l]{Decode time\\(5 features)}}
  & $\mu$ & Average decode time \\
  & $\sigma$ & Standard deviation of decode time \\
  & $\overline{\log t}$ & Average of $\log(\text{decode time})$ \\
  & $P_{95}$ & 95th percentile of decode time \\
  & $P_{99}$ & 99th percentile of decode time \\
\bottomrule
\end{tabular}
\end{table}

We arrived at this 14-feature vector (Table~\ref{tab:eval_fingerprint_features}) by  pruning an initial set of 44 features. The starting set included our 9-feature firing-count histogram, various summary statistics, and a one more histogram that chopped the range of decode times into 30 equal buckets and reported the fraction of shots landing in each. Ablating this set under cross-validation revealed two key lessons. First, the 30-bucket time histogram was redundant; it offered no more signal than a few simple percentiles, and its extra dimensions actively degraded our distance-based classifier with uninformative noise. Second, both remaining feature families are strictly necessary: dropping firing probabilities collapses accuracy at $d{=}7$, while removing timing statistics breaks it at $d{=}5$. The final 14 features form a minimal set required to succeed across both code distances.

\subsection{Classification}
Given a query fingerprint, the adversary predicts its source device by comparison against a \emph{reference bank}, which is a collection of fingerprints from a previously observed runs, each labeled by its device (\callout{4} in Figure~\ref{fig:eval_pipeline}). In this work we used a $k$-nearest-neighbor classifier. We chose this method because it is easy to interpret, and with a properly tuned $k$, it provides a stable baseline that is far less likely to overfit than more complex models. This model also matches the rolling nature of the attack; the adversary accumulates fingerprints over time and classifies each new run against the current set. Adding another fingerprint is free because $k$-NN simply stores it. A parametric model, such as a neural network would have to be retrained on the full dataset each time the reference bank grows and is much harder to interpret.

%% file: sections/experimental_setup.tex
\section{Experimental Setup}
\label{sec:setup}

\subsection{Devices and Calibration}
\label{sec:setup:devices}
We collect hardware data from three IBM Heron-R2 superconducting processors, \texttt{ibm\_marrakesh}, \texttt{ibm\_fez}, \texttt{ibm\_kingston}, accessed through Qiskit\cite{qiskit2024} Runtime. All three share the same Heron-R2 architecture and differ only in their physical realization and calibration. We do not control calibration: IBM recalibrates each device daily. Gate, readout, and idle error rates drift between calibrations. We treat this drift as part of the problem. Too much drift can make our fingerprints weaker as one device may look more like another device on a different day due to drift over time. We hold the mapping and the optimization level of our circuit fixed across all runs.

\subsection{Circuit: Repetition-Code Memory Experiment}
\label{sec:setup:circuit}

\begin{figure}[h]
    \centering
    \includegraphics[width=\linewidth]{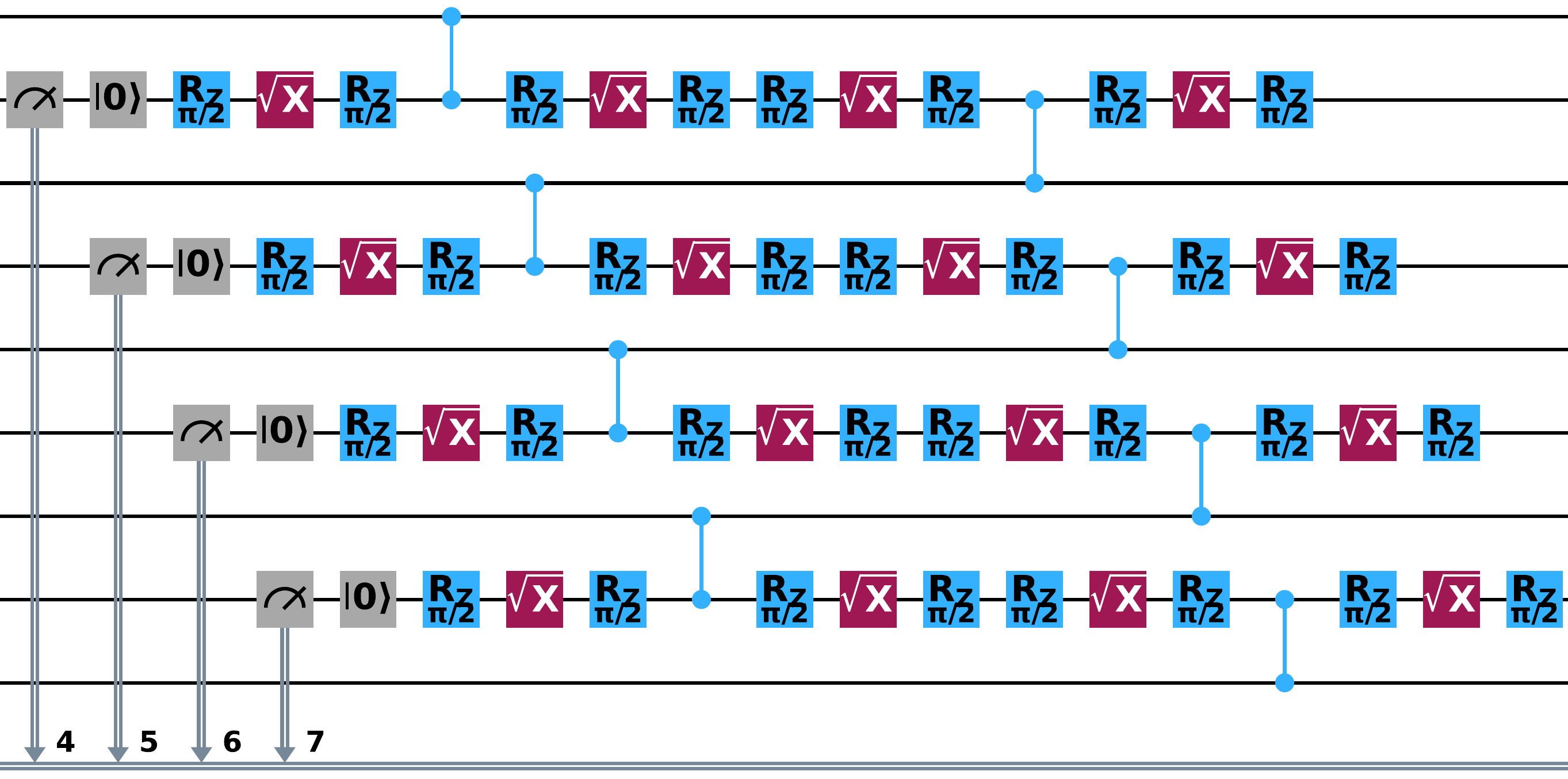}
    \caption{A segment of the $d{=}5$ repetition-code circuit on IBM Heron, transpiled to native gates ($R_Z$, $\sqrt{X}$, CZ). Mid-circuit measurements (meter symbols) extract syndrome bits and reset the ancilla qubits for the next round. These measurements generate the detector events that dictate the decoder's data-dependent latency.}
    \label{fig:circuit}
\end{figure}

On every device we run the same repetition-code memory experiment; a representative section of its transpiled circuit is shown in Figure~\ref{fig:circuit}. We use the repetition code because it exercises the exact mechanism our side channel targets, real syndrome-extraction rounds feeding a decoder on the critical path, while being shallow enough to run reliably and repeatedly on today's hardware across the full 68-day window. A distance-$d$ instance uses $d$ data qubits and $d-1$ ancilla qubits and runs at $d \in \{5,7\}$. Each shot prepares the data register in $|0\rangle$, performs two rounds of syndrome extraction, applies a transversal logical~$X$, performs three further rounds, and finally measures the data register. This gives us five syndrome-extraction rounds in total. Because the mid-circuit logical~$X$ flip the encoded state, the expected logical outcome of every shot is $|1\rangle$; a shot is counted as a logical error whenever the decoder's correction, applied to the measured observable, disagrees with that expected outcome. The circuit is the same across all devices, so any differences in the resulting timing distributions is attributable to the the hardware and not the workload.

\subsection{Timing Instrumentation}
\label{sec:setup:timer}

We execute and time all decodes locally on a single Linux workstation (Intel Xeon Silver 4509Y, Ubuntu) using raw syndromes from the IBM devices. To prevent operating-system jitter from obscuring sub-microsecond decodes, we pin the timing process to a dedicated, idle CPU core via \texttt{taskset}, which also mirrors how real-world fault-tolerant decoders are deployed as dedicated services.

Decoding graphs are derived from Stim~\cite{stim} detector error models. PyMatching's MWPM~\cite{pymatchingv2} consumes these directly, while for Union-Find, we convert them to parity-check matrices for the peeling decoder in the \texttt{ldpc} package~\cite{ldpc, delfosse2021union}. We time these compiled C/C++ decoders using nanosecond-resolution monotonic clocks. Because MWPM decodes are sub-microsecond, we avoid Python overhead by looping and timing (\texttt{clock\_gettime(CLOCK\_MONOTONIC\_RAW)}) entirely within a custom C extension. Union-Find decodes take tens of microseconds, making Python dispatch overhead ($\sim$100\,ns) negligible, so we time them directly in Python via \texttt{time.perf\_counter\_ns}. For both, we prime CPU caches with thousands of warm-up decodes before timed batches. Downstream analysis uses NumPy and scikit-learn~\cite{scikit-learn}.

Processing every device's syndromes through identical decoder instances on the same core is deliberate. It ensures the timing-to-firing-count mapping remains purely a property of the decoder software, preventing classical hardware differences (such as varying CPU clock speeds) from independently skewing latencies across different quantum devices.

\subsection{Dataset}
\label{sec:setup:dataset}
Our hardware corpus comprises 14 runs collected over 68 days across three devices. For each combination of run, device, and code distance, we execute 48,000 shots to compute the ground-truth logical error rate ($p_L$). We then process 16,000 of these shots through our timing harness, recording the per-shot decode time and, for validation only, the ground-truth detector-firing count.

\emph{Crucially, true firing counts are never used as attack features.} They serve exclusively to validate our pipeline. Every evaluation result in Section~\ref{sec:eval_hw} (including device fingerprints, inferred code distances, and $p_L$ estimates) is reconstructed entirely from decode timing. To infer the firing counts, we calibrate the mean log decode time for each count using a reference dataset. Each shot is then mapped to the nearest firing count within this log-time space. Consequently, the resulting firing-count features depend solely on decode timing, completely independent of the ground-truth syndromes.

We time two decoders: minimum-weight perfect matching (MWPM, via PyMatching) and Union-Find, running on an isolated CPU core. To ensure a fair comparison, the MWPM decoder uses a uniform error model ($p=10^{-3}$) rather than per-device calibrations. This means the decoder's internal matching graph assigns identical weights to all potential errors, deliberately denying the algorithm any prior knowledge of a specific chip's noise topology or localized defect rates. By applying this identical, uncalibrated decoder instance across all three devices, any difference in the resulting timing distributions is strictly attributable to the raw error patterns emitted by the hardware itself, not the decoder configuration.

%% file: sections/evaluation.tex
\section{Evaluation}
\label{sec:eval_hw}

Our evaluation is structured around three research questions. \textbf{RQ1:} Can a passive observer reconstruct the detector-firing count and, from its distribution, estimate the workload's logical error rate $p_L$ and empirical suppression factor $\Lambda$ using only decoder timing? \textbf{RQ2:} Can the code distance chosen  by the target workload be inferred from decode-time distribution? \textbf{RQ3:} Can specific physical hardware be reliably fingerprinted from decode-time distributions across days? We first answer these questions on real IBM Heron hardware (Sec.~\ref{sec:eval_hw}), where every measurement is anchored in a real Qiskit Runtime job. We then test our fingerprinting method with data from Google Willow machines (Sec.~\ref{sec:sim}) to confirm that the same attack mechanism carries into the below-threshold surface-code regime that many of today's available hardware cannot yet sustain \cite{google2025quantum}.

\subsection{Detector-Count Inference from Decode Times}
\label{sec:eval:detector_count}
\begin{figure}[h]
    \centering
    \includegraphics[width=\linewidth]{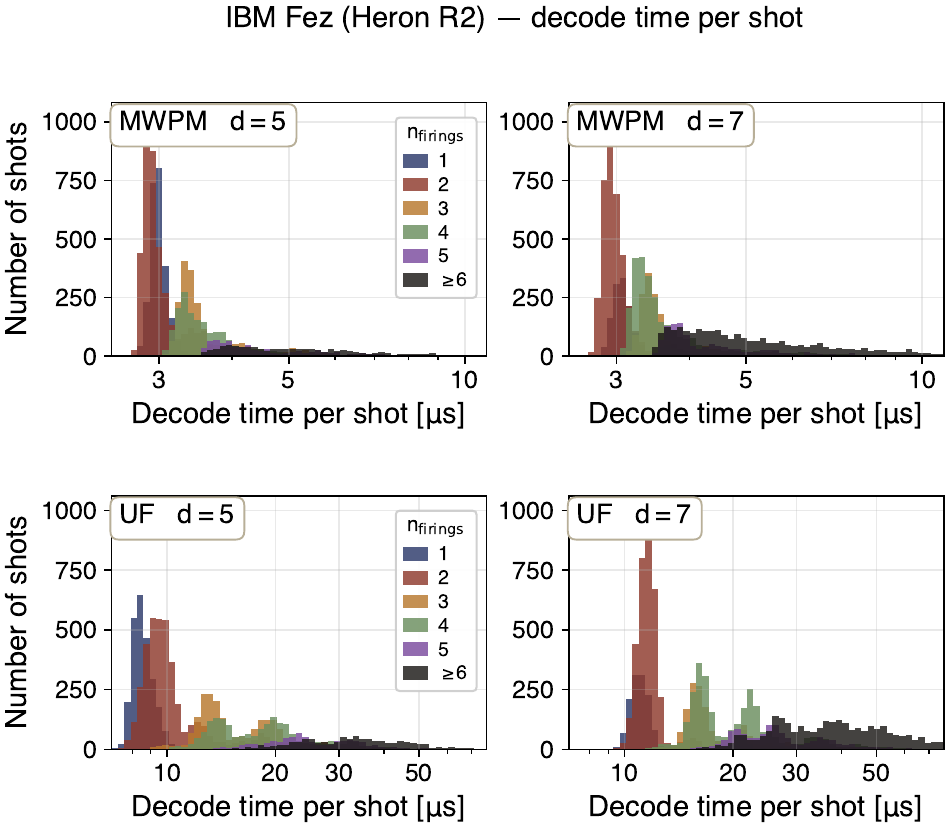}
    \caption{Per-shot decode time on IBM Fez (log scale; shots with at least one fired detector), grouped by detector-firing count $n_\mathrm{firings}$, for MWPM (top) and Union-Find (bottom) at $d{=}5$ (left) and $d{=}7$ (right). Union-Find's runtime is strongly data-dependent: each firing count occupies a distinct, progressively longer band, marching from $\sim$8 to 50,$\mu$s. MWPM is far less sensitive, compressing the same counts into a narrow $\sim$2.5 to 10,$\mu$s window where adjacent counts overlap more heavily. Trivial 0-firing shots, which decode almost instantly, are omitted for clarity.}
    \label{fig:eval_firings}
\end{figure}

We first investigate whether a passive observer can reconstruct the shot-by-shot detector-firing count, and subsequently estimate the logical error rate, relying entirely on decoder wall-clock time. The underlying mechanism is straightforward: because both MWPM and UF must expend more computational effort to resolve denser syndromes, per-shot latency naturally scales with the number of fired detectors. Figure~\ref{fig:eval_firings} illustrates this effect on a single IBM Heron device (\texttt{ibm\_fez}), plotting the per-shot decode-time distributions for both decoders at code distances 5 and 7.

\begin{figure}[h]
    \centering
    \includegraphics[width=\linewidth]{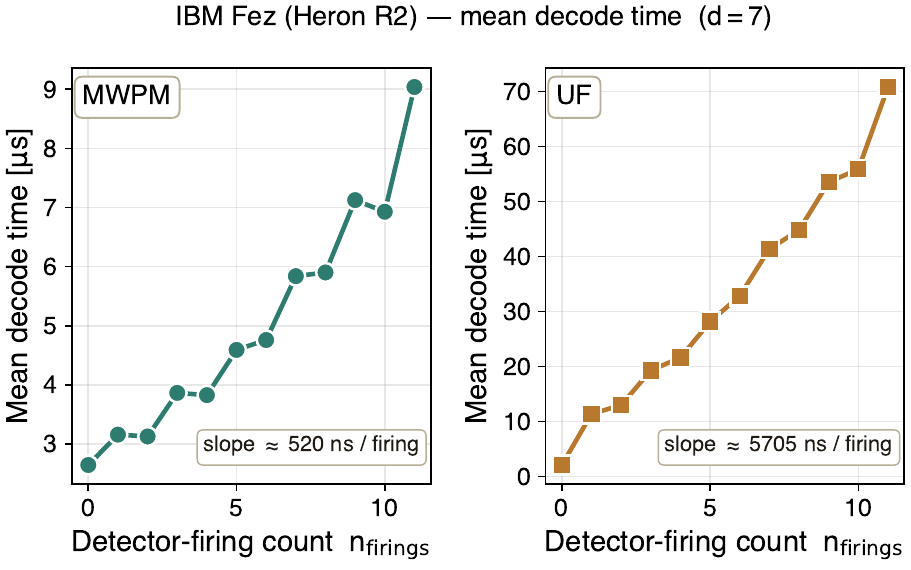}
\caption{Mean per-shot decode time on IBM Fez at $d{=}7$ for MWPM (left) and Union-Find (right), each on its own scale. In both decoders the mean latency grows linearly with the fired-detector count, confirming that decode time is data-dependent regardless of which decoder is used. The panels use independent scales; we do not compare absolute latencies across decoders, as UF is a reference Python implementation and MWPM is PyMatching.}
    \label{fig:eval_firings_meansx}
\end{figure}

We ran 16{,}000 timed shots per code distance in a single run on \texttt{ibm\_fez} (Heron R2) and plotted the per-shot decode-time distribution grouped by firing count in Figure~\ref{fig:eval_firings_meansx}. For Union-Find, we see distinct, clearly separated distributions at each firing count for both $d{=}5$ and $d{=}7$; for MWPM, the distributions overlap noticeably more. Both decoders nevertheless confirm the underlying mechanism: mean execution time scales consistently with the number of fired detectors.

Neither decoder is timing-flat by design: both solve a syndrome-dependent optimization problem, so their runtime is a function of the input, not a fixed cost. Union-Find pays that cost per fired detector directly, which is why its slope is steep. MWPM amortizes most of its work into precomputation and then does only a small per-firing update, which flattens its slope by an order of magnitude but does not eliminate it. Both decoders leak the firing count through time, but UF just leaks it louder.

\subsection{Logical Error Rate}
Now that we have shown the detector-firing count can be reliably extracted through timing alone, we address our next core question: can this underlying distribution be used to derive the logical error rate?

To evaluate this, we pooled data from 14 runs collected over a 68-day window (May 4-July 10, 2026) on the three IBM Heron devices (\texttt{ibm\_marrakesh}, \texttt{ibm\_fez}, and \texttt{ibm\_kingston}). We grouped this data into discrete cells defined by the specific run, device, and code distance. For each cell, we first established a ground-truth logical error rate ($p_L$) by decoding all 48,000 shots. Because we initialized the data qubits into a known state and applied just a logical X in our circuit, the expected final logical outcome was fixed. We could compute the logical error rate directly by comparing the decoder's prediction for each shot against the circuit's \emph{expected} logical outcome, and simply counting the fraction of shots where the applied correction failed to restore the correct logical state. Next, to simulate what a passive observer would see, we estimated the mean detector-firing count $\overline{n_\mathrm{firings}}$ using \emph{only} the decode times of those same shots.

To estimate $p_L$ directly from decoder timing we used a simple linear regression model. To fit a regression model we had to transform our data. Across our cells, $p_L$ grows non-linearly with $\overline{n_\mathrm{firings}}$, so a plain line through $(\overline{n_\mathrm{firings}}, p_L)$ would fit the high-$p_L$ cells and badly miss the low-$p_L$ ones. Taking the logarithm of $p_L$ removes that mismatch: on a log axis, the power-law dependence becomes a straight line, and ordinary least-squares can do the rest. We therefore fit one per-distance log-linear model,

$$ \log_{10} p_L = a \, \overline{n_\mathrm{firings}} + b. $$

The two coefficients have a direct interpretation: each additional fired detector (on average) adds $a$ to $\log_{10} p_L$, i.e., multiplies the underlying logical error rate by a factor of $10^a$.

\begin{figure}[h]
    \centering
    \includegraphics[width=\linewidth]{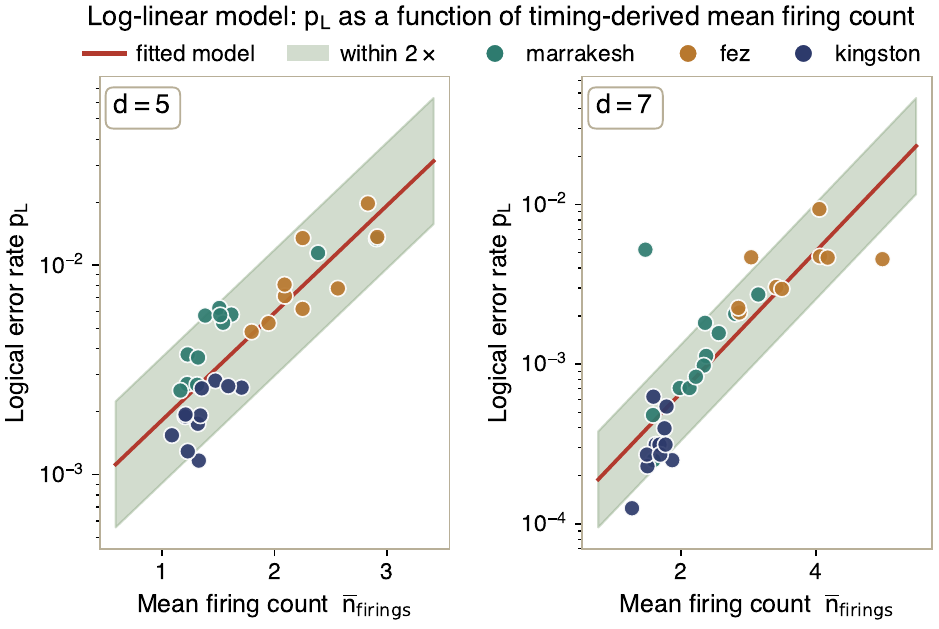}
    \caption{Log-linear models predicting the logical error rate ($p_L$) from the estimated mean firing count ($\overline{n}_\mathrm{firings}$) for $d{=}5$ (left) and $d{=}7$ (right). Each point represents a single run. The red line shows the regression fit ($\log_{10} p_L = 0.513\,\overline{n}_\mathrm{firings} - 3.25$ for $d{=}5$; $\log_{10} p_L = 0.443\,\overline{n}_\mathrm{firings} - 4.07$ for $d{=}7$). The green band highlights predictions within a factor of two of the truth ($\pm 2\times$), successfully capturing 94\,\% of $d{=}5$ and 84\,\% of $d{=}7$.}
    \label{fig:eval_pl_model}
\end{figure}

Figure~\ref{fig:eval_pl_model} shows the model itself, fit separately for each code distance. Each point represents a single run on a specific device; the horizontal position is the  mean detector-firing count ($\overline{n}_\mathrm{firings}$) for that run based on timing, and the vertical position is the ground-truth logical error rate $p_L$ we measured from the same shots on hardware. The color of each point identifies which physical processor ran the circuit. The red line is our fitted log-linear model; the shaded region around it is the factor-of-two envelope. Any point that falls inside this band is a run whose $p_L$ our model predicts to within a factor of two of the true value \footnote{Accuracy figures are strictly out-of-sample, computed using leave-one-run-out cross-validation. The quoted coefficients provide a descriptive summary of the full dataset.}

These results demonstrate that decoder timing can approximate the logical error rate. To gauge how much the leak actually buys an attacker, we compare it against the best estimate available without timing. Across our dataset $p_L$ spans nearly two orders of magnitude, so with no side channel an attacker's best guess is the dataset-wide median, which falls within a factor of two of the true $p_L$ only 59\% of the time at $d{=}5$ and 25\% at $d{=}7$. The timing side channel raises this to 94\% and 84\% respectively, turning a coarse prior into a factor-of-two estimate of how well a workload is being protected. This predictive power remains robust despite 68 days of drifting hardware calibration across three different processors, showing that a passive observer armed only with wall-clock times can approximate the logical error rate.

\subsection{Code-Distance Discrimination}
\label{sec:eval:distance}

We now investigate whether this same timing signal reveals the underlying code distance. Because distance is a primary parameter a FTQC uses to trade physical qubit overhead for logical fidelity, exposing it tells a passive observer exactly how defensively the operator is running the workload. We demonstrate that an attacker can reliably extract this configuration parameter, provided they observe a sufficient number of executions.

\begin{figure}[t]
    \centering
    \includegraphics[width=\linewidth]{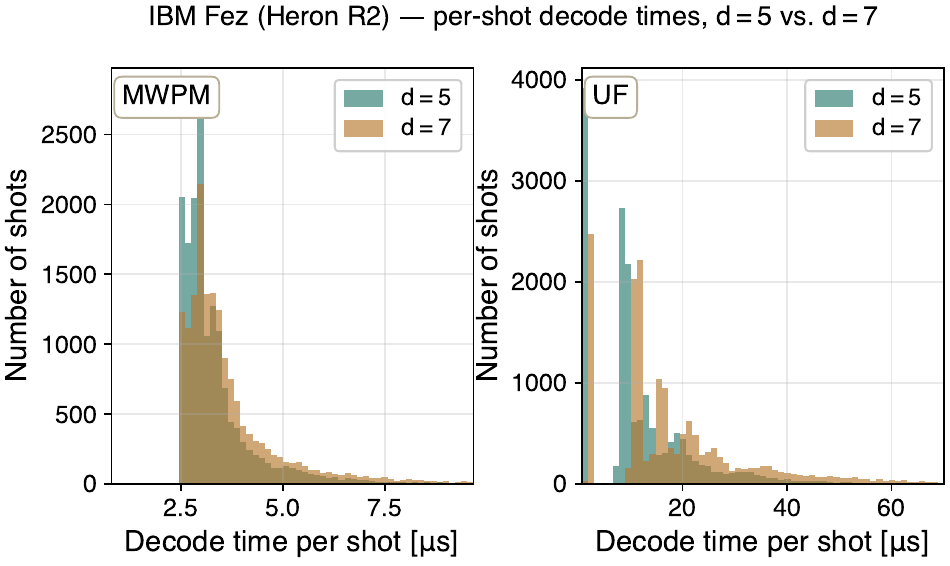}
    \caption{Per-shot decode-time distributions on IBM Fez for $d{=}5$ (teal) and $d{=}7$ (amber), shown separately for MWPM (left) and Union-Find (right). Increasing the code distance shifts the distributions toward noticeably longer execution times, allowing a passive observer to distinguish the two parameters from a batch of shots.}
    \label{fig:eval_firings_means}
\end{figure}

To determine if distance discrimination is possible, we first analyzed the aggregate decode-time data to see if the underlying distributions were separable. As Figure~\ref{fig:eval_firings_means} illustrates, increasing the code distance pushes the timing distribution toward longer latencies. To confirm this shift statistically, we ran a two-sample Kolmogorov-Smirnov (KS) test on 16{,}000 shots per distance from a single Fez run. We obtained a KS statistic of $D=0.178$ for MWPM ($p<10^{-221}$) and $D=0.401$ for Union-Find ($p<10^{-300}$), confirming that the two code distances produce statistically distinguishable decode-time distributions.

Having established that the distributions are distinct, we next sought to answer a practical threat-modeling question: exactly how many shots does an attacker need to observe to reliably extract the distance? We first measured the mean-shift effect sizes. These are small for MWPM (Cohen's $d = 0.10$) and moderate for Union-Find (Cohen's $d = 0.56$), so individual shots overlap too much to provide a reliable signal, especially for MWPM, whose mean latency barely shifts with distance. Consequently, the attacker must observe a \emph{batch} of shots, combined with a reasonably fresh noise baseline (Section~\ref{sec:eval:baselines}), to overcome this variance. This also explains why MWPM needs more shots than Union-Find to reach the same accuracy: its smaller per-shot mean shift takes longer to average out.

\begin{figure}[h]
    \centering
    \includegraphics[width=0.8\linewidth]{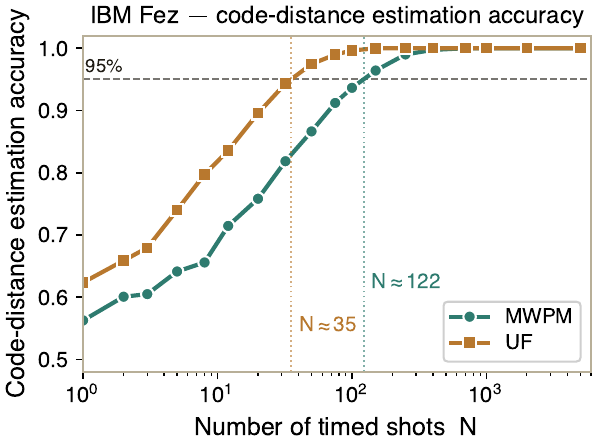}
    \caption{Code-distance estimation accuracy on IBM Fez as a function of the number of timed shots $N$ observed. Under a same-run reference baseline, the batch mean decode time identifies the code distance in 95\,\% of runs after $N\!\approx\!35$ shots for Union-Find and $N\!\approx\!122$ shots for MWPM, exceeding 99\,\% by $N\!\approx\!250$.}
    \label{fig:eval_how_many_shots}
\end{figure}

Figure~\ref{fig:eval_how_many_shots} quantifies the effectiveness of this batched approach by plotting the attacker's accuracy as a function of the number of observed shots ($N$). Assuming a same-run reference baseline, calculating the batch mean decode time correctly identifies the code distance in 95\%\footnote{Accuracy is the fraction of 4{,}000 Monte-Carlo trials in which the attacker's guess matched the true distance; an accuracy of 50\% corresponds to guessing code distance at random.} of runs after observing just $N\!\approx\!35$ shots for Union-Find and $N\!\approx\!122$ shots for MWPM\footnote{We report these shot counts as approximate because we only measured accuracy at specific intervals. The exact thresholds are estimated by interpolating between those measured points.}. By $N\!\approx\!250$ observed shots, the attacker's accuracy exceeds 99\%.

\subsection{Baselines}
\label{sec:eval:baselines}
Given that underlying quantum hardware noise drifts continuously, a critical question arises: can an attacker still extract useful configuration details, such as the target's code distance and number of detector firings when the baseline is a moving target? If so, how frequently must they recalibrate to maintain a reliable information leak?

\paragraph{\textbf{Extracting firing counts relies on a single, universal baseline}}
The mean decode time for a given number of fired detectors is a property of the classical decoder software, essentially independent of the QPU that generated the syndrome. Across all three Heron~R2 devices (\texttt{ibm\_marrakesh}, \texttt{ibm\_fez}, and \texttt{ibm\_kingston}), the mean execution time for a given code distance $d$ and $n_\mathrm{firings}$ cell is nearly identical, varying by only a few percent (median under 4\%, at most $\sim$8\%). Over a 68-day window on \texttt{ibm\_fez}, the same mean latency drifts by a median of 8 to 11\%, and more (up to $\sim$25\%) on occasional recalibration days. A timing baseline therefore transfers well across devices and time, so an attacker builds the timing-to-firing-count profile once and reuses it. Because the decode-time distributions of adjacent firing counts overlap, individual shots cannot always be assigned an exact firing count; but a batch of shots recovers the firing-count distribution accurately enough to drive both the device fingerprint and the $p_L$ estimate.

\paragraph{\textbf{Extracting code distance requires periodic, device-specific baselines}}
While the decoder's timing is stable, the actual number of detectors that fire depends heavily on the QPU's underlying physical error rate, which drifts daily and varies across chips. A noisier run shifts the firing distributions for both $d{=}5$ and $d{=}7$ toward higher counts. The code distance itself manifests as a systematic offset, $d{=}7$ consistently produces slightly more firings than $d{=}5$ at the exact same noise level, but the absolute placement of these distributions on the count axis slides as hardware noise fluctuates. Because identifying the code distance requires tracking this moving target, the attacker cannot rely on a universal baseline. Instead, our threat model conservatively assumes the attacker must periodically refresh their noise profile by capturing at most one known-distance per device near calibration.

\subsection{Hardware Fingerprinting}
\label{sec:eval:hf}
The final question is whether the decode-time side channel also leaks the identity of the underlying physical hardware. Extracting the specific device identity gives an attacker information that could be used to enable more sophisticated exploits. Most notably, device fingerprinting allows an adversary to confirm co-residency with a high-value target, a prerequisite for launching cross-tenant attacks.

\begin{figure}[h]
    \centering
    \includegraphics[width=\linewidth]{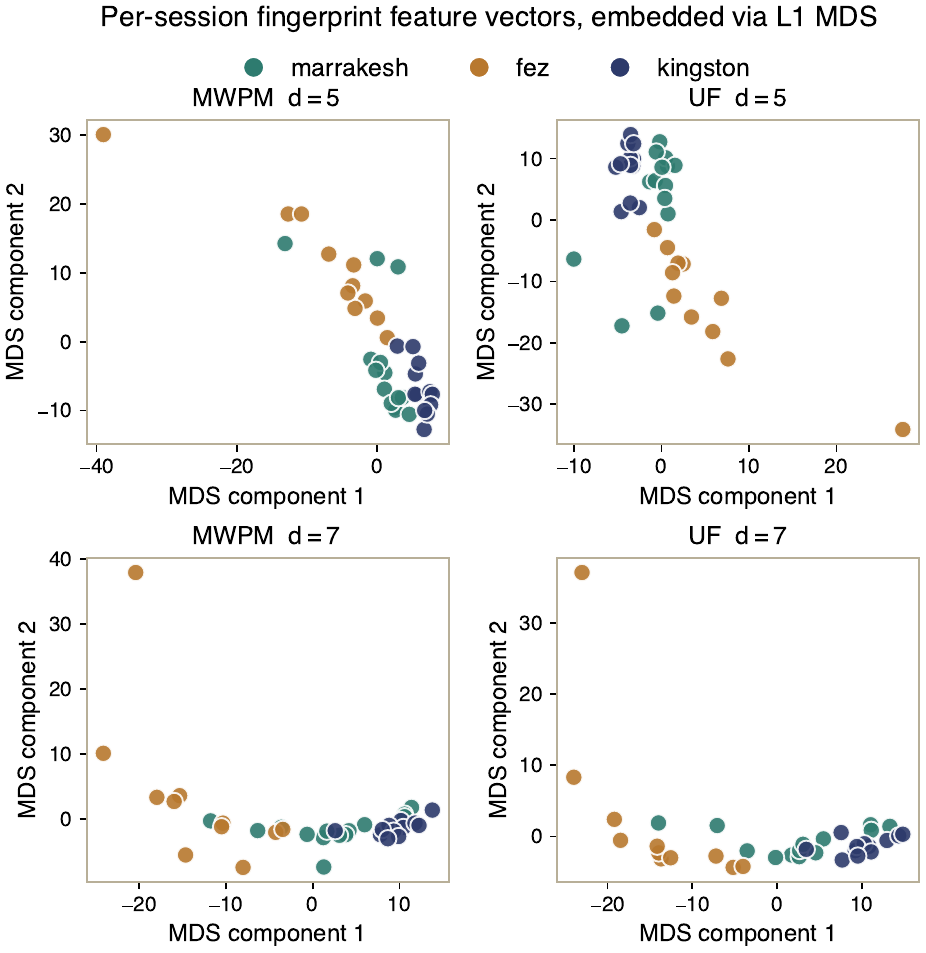}
     \caption{A 2D Multidimensional Scaling (MDS) projection of the per-session fingerprint feature vectors (Table~\ref{tab:eval_fingerprint_features}) Each point represents a single run mapped from the 14-dimensional feature space used by our classifier (nine firing-count probabilities and five decode-time tail statistics). The layout is based on the Manhattan (L1) distance between standardized vectors, meaning sessions with similar timing profiles naturally cluster together. Panels are organized by code distance (rows) and decoder (columns).}
    \label{fig:eval_mds}
\end{figure}

To visualize whether these fingerprints separate by device, we take every run's 14-dimensional feature vector, standardize it, and compute the pairwise Manhattan (L1) distance between all runs; the same distance metric our classifier uses. We then apply classical multidimensional scaling (MDS), which places each run's feature vector as a point in 2D space so that Euclidean distances between points reproduce the L1 distances as closely as possible; runs that have similar fingerprints land closer together. Figure~\ref{fig:eval_mds} colors the resulting layout by device. Runs from the same physical processor form visually distinct clusters in every (distance, decoder) panel. We stress that this separation is a property of the full feature space, MDS is a lossy 2D projection used only for visualization, and the classifier below operates on the raw vectors.

\subsubsection{KS Testing}
To quantify how distinct these fingerprints are, we test the distributions using a pairwise KS test. In Section~\ref{sec:eval:distance}, we used the KS test to verify the statistical separation between different code distances on a single device. Here, we apply the exact same metric for fingerprinting: holding the code distance constant, we use the KS test to determine whether the decode-time distributions significantly diverge across different physical machines.

We ran a two-sample KS test between every pair of devices, at each code distance and for each decoder, and recorded the KS statistic $D$. $D$ is the maximum distance between the cumulative distribution functions (CDFs) of two decode-time distributions: a value of $0$ means the distributions are indistinguishable, and a larger value means they are easier to tell apart. Table~\ref{tab:eval_ks} reports the results. Every pair is separated with overwhelming significance ($p < 10^{-100}$), and the separation grows with code distance. The most distinct cell, fez vs.\ kingston at $d{=}7$ under MWPM, reaches $D=0.425$.

\begin{table}[t]
\centering
\caption{Pairwise Kolmogorov--Smirnov statistics between the three IBM
Heron devices, (all $p < 10^{-100}$). Larger values indicate the timing distributions are
more distinct and easier to distinguish.}
\label{tab:eval_ks}
\begin{tabular}{lcccc}
\toprule
& \multicolumn{2}{c}{MWPM} & \multicolumn{2}{c}{Union-Find} \\
\cmidrule(lr){2-3} \cmidrule(lr){4-5}
Pair                    & $d{=}5$ & $d{=}7$ & $d{=}5$ & $d{=}7$ \\
\midrule
marrakesh vs.\ fez      & 0.246  & 0.300  & 0.233  & 0.288 \\
marrakesh vs.\ kingston & 0.050  & 0.131  & 0.055  & 0.133 \\
fez vs.\ kingston       & 0.295  & 0.425  & 0.280  & 0.407 \\
\bottomrule
\end{tabular}
\end{table}

\subsubsection{Device Identification}

\begin{table}[t]
\centering
\caption{Device identification accuracy using only decoder timing. Each run is summarized by the 14-dimensional feature vector (fingerprint) and evaluated using a $k$-nearest-neighbor classifier ($k{=}3$). Random guessing yields 33.3\,\%.}
\label{tab:eval_device_classifier}
\begin{tabular}{llcccc}
\toprule
Decoder & Distance & Overall & Marrakesh & Fez & Kingston \\
\midrule
MWPM        & $d{=}5$ & 86.5\% & 79\% & 91\%  & 92\%  \\
MWPM        & $d{=}7$ & 81.1\% & 71\% & 82\%  & 92\%  \\
Union-Find  & $d{=}5$ & 89.2\% & 79\% & 91\%  & 100\% \\
Union-Find  & $d{=}7$ & 89.2\% & 79\% & 100\% & 92\%  \\
\bottomrule
\end{tabular}
\end{table}

The KS test confirms that there is a statistically significant difference between the decode-time distributions of each machine, but an attacker must actually assign an unknown run to the correct device. Given a newly collected fingerprint from a run, our $k$-nearest-neighbor classifier assigns it the majority device label among its three nearest fingerprints in space.

Table~\ref{tab:eval_device_classifier} reports the results. Timing alone identifies the correct device in 81 to 89\% of runs for every decoder and distance, well above the
one-in-three level a random guess would reach, with Union-Find best at 89.2\%. \texttt{ibm\_fez} and \texttt{ibm\_kingston} are identified reliably (usually above 90\%), while \texttt{ibm\_marrakesh} is the hardest (71 to 79\%) and accounts for most of the residual error.

The per-detector firing rates explain why. The repetition code lays its qubits out in a line, so each detector's firing rate maps where errors occur along the chain. Each
device has a stable spatial signature: \texttt{ibm\_kingston} has one hot chain position in all 12 of its runs and \texttt{ibm\_fez} is consistently hot at both ends, whereas
\texttt{ibm\_marrakesh} has no stable hotspot and the flattest profile. Identifiability tracks this spatial pattern, not the average error rate: \texttt{ibm\_kingston} has the lowest mean firing rate yet is identified reliably, because a fixed bad qubit is a durable, localized signature. The fingerprint captures where a device's noise lives, not just how much.

\subsection{Generalization to Surface Codes: Google Willow}
\label{sec:sim}

The timing side channel is not confined to repetition codes or to a single vendor: it works on real, below-threshold surface-code hardware. Future fault-tolerant workloads may run surface codes \cite{google2025quantum}, so we test our pipeline on public data from Google's 105-qubit Willow processor, the first device to demonstrate quantum error correction below the surface-code threshold \cite{google2025quantum}. Its released memory-experiment dataset contains the actual per-shot detection events measured on the chip. Because surface codes with larger distances fire significantly more detectors than repetitions codes, saturating our original fixed timing-derived firing-count bins, we generalize the fingerprint (Table~\ref{tab:eval_fingerprint_features}) for large-scale machines by relying entirely on the decode-time stats: the mean ($\overline{\log t}$), standard deviation ($s_{\log t}$), and percentiles ($P_5, P_{10}, \dots, P_{99}$) of the log decode times.

The Willow processor executes multiple distance-3 surface-code patches at varying physical locations across the chip, with each patch utilizing a distinct set of qubits exposed to local noise variations. Constrained to a single physical chip, we cannot compare entirely separate devices. Instead, we treat each of the 9 patches as an independent ``sub-device'' to evaluate whether decoder timing alone can distinguish between them. Furthermore, unlike our Heron evaluations which span multiple calibration cycles, this dataset represents a single-day experimental snapshot.

\begin{figure}[h]
\centering
\includegraphics[width=0.7\linewidth]{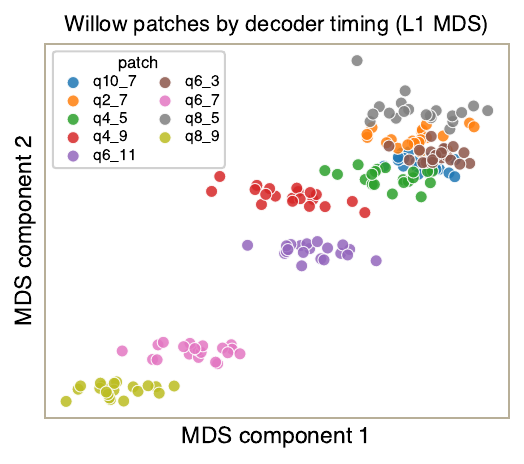}
\caption{Per-patch decode-time fingerprints on Google Willow, embedded in 2D via L1 MDS. Each point is a sub-sample of one of the 9 distance-3 surface-code patches; points cluster by patch, so decoder timing alone separates the physical patches. Measured on a dedicated, core-pinned Linux core.}
\label{fig:willow}
\end{figure}

For every patch, we decode its physical detection events ($50{,}000$ shots, $50$ syndrome rounds) using a \emph{single, fixed} MWPM decoder initialized once with a uniform SI1000 noise model. This guarantees that any variation in decode latency stems purely from the patch's underlying syndrome statistics rather than decoder configuration. Consistent with our prior methodology, all decodes are executed and timed on a dedicated, core-pinned Linux CPU to isolate the signal.

Decoder timing alone identifies which of the 9 patches a run came from with 81\% accuracy (random guessing is 11\%). Figure~\ref{fig:willow} visualizes this. We split each patch's 50{,}000 shots into 20 equal batches of 2{,}500 shots and compute one decode-time fingerprint from each batch. \emph{Each point in the figure is one such fingerprint}, embedded in 2D via L1 MDS, so every patch contributes 20 points (180 in total across the 9 patches). The points cluster cleanly by patch: each patch's local noise imprints a characteristic distribution of firing counts, and hence of decode times, that the timing recovers. The pattern mirrors our results on the IBM Heron machines. Patches with distinctive noise produce decode-time distributions that stand apart from the rest, forming tight, well-separated clusters that are identified almost perfectly; patches whose noise resembles that of their neighbors produce overlapping distributions and are the hardest to place, just as the least distinctive Heron device, \texttt{ibm\_marrakesh}, was the hardest to identify.

\begin{table}[h]
\centering
\caption{Intra-chip patch fingerprinting on Google Willow across code distances. The $d{=}7$ region holds a single patch, so no patch-identification task exists there.}
\label{tab:willow}
\begin{tabular}{cccccc}
\toprule
Distance & Patches & Detectors & Mean decode & Accuracy & Chance \\
\midrule
$d{=}3$ & 9 & 400   & $10.4\,\mu$s & 81\% & 11\% \\
$d{=}5$ & 4 & 1{,}200 & $40.7\,\mu$s & 82\% & 25\% \\
$d{=}7$ & 1 & 2{,}400 & $92.3\,\mu$s & ---  & --- \\
\bottomrule
\end{tabular}
\end{table}

Scaling to higher code distances reveals that mean decode time directly leaks the distance itself, jumping from $10.4\,\mu$s at $d{=}3$, to $40.7\,\mu$s at $d{=}5$, and $92.3\,\mu$s at $d{=}7$ (Table~\ref{tab:willow}). All decoding was done with MWPM \cite{pymatchingv2}.  Because these latency distributions never overlap, a passive observer can extract the code distance purely from timing, mirroring our Heron results. Furthermore, intra-chip patch fingerprinting persists at scale: decoder timing identifies the 9 $d{=}3$ patches at 81\% and the 4 $d{=}5$ patches at 82\%. While the single $d{=}7$ patch precludes an identification task, its extensive latency confirms the timing signal only amplifies as codes grow.

%% file: sections/discussion.tex
\section{Discussion}
\label{sec:discussion}
Our threat model assumes an observer can measure per-shot latencies from a data-dependent decoder. Consequently, defenses must either eliminate the data-dependence or prevent latency observation entirely. We discuss these mitigations and outline our study's limitations.

\paragraph{Constant-Time Decoding}
\label{sec:disc:constant_time}
The most direct defense is constant-time decoding, ensuring every shot takes identical wall-clock time regardless of the syndrome. Since standard MWPM and Union-Find algorithms are inherently data-dependent, this requires structural changes. The first approach is latency padding: delaying the output until a worst-case fixed time budget elapses. This eliminates leakage but throttles throughput to worst-case speeds. The second approach uses fixed-latency architectures, such as fully pipelined systolic hardware decoders, which perform a constant volume of work per cycle. Both methods close the channel by construction but introduce steep trade-offs between security, hardware complexity, and real-time processing rates.

\paragraph{Obscuring Decode Latency}
\label{sec:disc:batching}
Alternatively, defenders can attempt to deny clean per-shot measurements without modifying the decoder. Shot batching obfuscates granular per-shot latencies by exposing only aggregate execution times, though this coarse timing can still leak the mean detector-firing rate unless combined with execution padding. Injecting random delays into decoding operations is a much weaker defense; an adversary can simply collect more shots to average away the noise, merely increasing sample complexity rather than closing the channel. Ultimately, strict architectural isolation, ensuring no co-located or unprivileged process can observe execution times, is required to fundamentally prevent these timing attacks.

\paragraph{Limitations}
\label{sec:disc:limits}
Our study has several limitations. First, we deployed identical circuits across all tested devices to control for baseline variations, leaving the disentanglement of workload-specific effects from device-specific noise to future work. Second, our empirical evaluation is bounded to three IBM Heron devices and two code distances; we do not evaluate how classification accuracy scales as the pool of candidate devices grows. Finally, our evaluation covers the two most widely deployed decoder families, minimum-weight perfect matching and Union-Find. We do not evaluate iterative decoders for qLDPC codes, such as BP+OSD, whose runtimes are even more strongly data-dependent. We leave a full characterization to future work.

%% file: sections/related_work.tex
\section{Related Work}
\label{sec:related}

\subsection{Side-Channel Attacks on Quantum Computers}
Physical side channels in quantum systems typically target the control stack. Xu et al.~\cite{xu2023power} and Erata et al.~\cite{erata2024power} reconstruct circuits using microwave power traces, while Bell et al.~\cite{bell2022reconstructing} recover parameters via interleaved dummy circuits. These require physical probing or active co-execution. Closest to our work, \emph{Quantum Leak}~\cite{lu2024quantumleak} uses the total circuit execution time on IBM's cloud to distinguish backends. However, that coarse signal depends on circuit depth and gate count in NISQ workloads. Our channel is fundamentally different: we target the \emph{per-shot} latency of the classical decoder, a component unique to fault-tolerant workloads. To our knowledge, we are the first to identify the decoder as a side-channel source. Other attacks exploit crosstalk in multi-programmed NISQ devices~\cite{ashsaki2020crosstalk, deshpande2022antivirus, mi2022reset}, which requires active co-execution, whereas our timing observer is strictly passive.

\subsection{Quantum Hardware Fingerprinting}
Prior fingerprinting identifies devices using directly readable information. Smith et al.~\cite{smith2022fast} use published qubit frequencies, noting they remain stable while error rates drift. Intrinsic processor features, such as qubit frequency, were also used as a basis for a quantum physical unclonable function~\cite{Tonekaboni2025}. We also note that quantum algorithms have been developed for checking the equivalence of files, a task also known as fingerprinting~\cite{anschuetz2026supercheq}.

Most closely related, Mutolo et al.~\cite{mutolo2025syndromes} fingerprint backends with 99\% accuracy using raw \emph{error-syndrome bit strings}. Other methods rely on error evolution~\cite{quanguard}, localized noise~\cite{localized}, locality signatures~\cite{allen2021device}, or quantum PUFs~\cite{phalak2021qpuf, skoric2017puf}. All these schemes require access to direct hardware observables like calibration data, syndromes, or specific challenge circuits. In contrast, our fingerprint relies \emph{exclusively} on classical decoder timing. The attacker never observes the syndrome bits, calibration data, or the circuit itself, yet can still identify the underlying physical device up to 89\% of the time. We demonstrate that device identity leaks even when the underlying syndromes are completely hidden.

\subsection{Decoder Latency and Real-Time Decoding}
The systems community extensively studies decoder latency for performance rather than security. Software decoders like PyMatching's MWPM~\cite{pymatchingv2} and Union-Find~\cite{delfosse2021union} exhibit syndrome-dependent runtimes. Hardware architectures, such as Micro Blossom~\cite{wu2025microblossom} and FPGA-based Union-Find~\cite{liyanage2024fpga}, are engineered specifically to bound worst-case latency. This decode-time variance is a recognized obstacle to real-time decoding, prompting solutions like elastic decoders~\cite{maurya2026elastic} to manage backlogs caused by difficult syndromes. Our contribution demonstrates that this well-known performance bottleneck also acts as a highly exploitable security side channel, and we formally characterize the information it leaks.

%% file: sections/conclusion.tex
\section{Conclusion}
\label{sec:conclusion}
We demonstrated that the wall-clock execution time of a fault-tolerant quantum decoder constitutes an exploitable side channel. Because MWPM and Union-Find runtimes scale with syndrome density, a passive observer can recover sensitive workload information from per-shot decode latencies alone. Using 68 days of data from three IBM Heron processors, we showed that timing alone estimates the logical error rate $p_L$ to within a factor of two on 84--94\% of runs, infers code distance with 95\% accuracy from as few as 50 timed shots, and fingerprints the physical processor with up to 89\% accuracy. Our Willow-inspired experiments confirm this fingerprinting generalizes to surface codes.

%% file: sections/acknowledgment.tex
\section*{Acknowledgment}
We acknowledge the use of IBM Quantum services for this work~\cite{ibmquantum}. The views expressed are those of the authors, and do not reflect the official policy or position of IBM or the IBM Quantum team.